\documentclass[aps,preprintnumbers,showpacs,twocolumn]{revtex4-1}

\usepackage{graphicx}
\usepackage{dcolumn}
\usepackage{bm}
\usepackage{amssymb}

\begin{document}

\preprint{}

\title{Global synchronization of bursting neurons in clustered networks}

\author{C. A. S. Batista,$^1$ R. V. Nunes,$^1$ A. M. Batista,
$^2$ R. L. Viana,$^3$, S. R. Lopes$^3$ and T. Pereira$^4$}
\affiliation{$^1$Departamento de F\'isica, Universidade Estadual de Ponta 
Grossa, 84030-900, Ponta Grossa, Paran\'a, Brazil.\\ 
$^2$Departamento de Matem\'atica e Estat\'istica, Universidade Estadual
de Ponta Grossa, 84030-900, Ponta Grossa, Paran\'a, Brazil\\
$^3$Departamento de F\'isica, Universidade Federal do Paran\'a, 81531-990,
Curitiba, Paran\'a, Brazil.\\
$^4$Centro de Matem\'atica, Computa\c c\~ao e Cogni\c c\~ao, Universidade 
Federal do ABC, 09210-170, Santo Andr\'e, S\~ao Paulo, Brazil.}
 
\date{\today}

\begin{abstract}
We investigate the collective dynamics of bursting neurons on clustered 
network. The clustered network is composed of subnetworks each presenting a 
small-world property, and in a given subnetwork each neuron has a probability 
to be connected to the other subnetworks. We give bounds for the critical 
coupling strength to obtain global burst synchronization in terms of the 
network structure, i.e., intracluster and intercluster probabilities 
connections. As the heterogeneity in the network is reduced the network global 
synchronization is improved. We show that the transitions to global synchrony 
may be abrupt or smooth depending on the intercluster probability.
\end{abstract}


\maketitle

\section{Introduction}

The human brain is a complex network consisting of approximately 
$10^{11}$ neurons, linked together by $10^{14}$ to $10^{15}$ connections, 
amounting to $10^4$ synap\-ses per neuron \cite{buzsaki}. There are
evidences that synchronization may be related  to a series of processes in 
the brain. Synchronized rhythms have been observed  in electroencephalograph
recordings of electrical brain activity, and are thought to be an important 
mechanism for neural information processing \cite{salinas}. 
Experimental observations reveals  synchronized oscillations of neuron networks 
in response to sensory stimuli in a variety of brain areas  
\cite{rodriguez, fries}. Such synchronized rhythms reflect the hierarchical organization of the brain, and  occur over a wide range of both, spatial and
temporal scales \cite{stam}. 

Complex network such as the brain are typically composed by subnetworks
 that are sparsely linked, leading to the appearance of clusters. 
Typically, models for brain activity deals with clustered
network with small-world architecture \cite{hagmann}.
 
Certain neurons in the brain exhibit burst activity: bursts of multiple spikes 
followed by a rest state hyperpolarization. These bursting neurons are important 
in different aspects of brain function such as movement control and cognition 
\cite{schultz, freeman, grace}.  

If the neurons possess two distinct time-scales such as spiking and bursting, 
the bursts (slow time-scale) tend to synchronize at smaller synaptic strengths 
\cite{dhamala, pereira1}.  We
consider two or more neurons to be bursting in a synchronized manner if they
start a given burst at  nearly the same time, even though the fast spiking may 
not be synchronized. We want to analyze the collective dynamics the a 
clustered network of bursting neurons and give bounds for the smallest coupling 
strength needed to achieve global synchronization on the bursting scale.

We investigate bursting dynamics on excitatory clustered networks. Close 
to the global synchronized state the phase dynamics of the burst may be 
described,  to first order approximation, by a Kuramoto-like model.  Further
investigations in the Kuramoto model allows us to determine the critical 
coupling strength to obtain burst global synchronization in terms of the 
probabilities of intracluster and intercluster connections. 

We show that the route to a global synchronized state reveals to distinct types 
of transitions as a function of the interclusters probabilities. If the intercluster
probability is small enough then the transition to global synchronization
is smooth, as opposed to higher values of the intercluster probability. In the 
latter case the network may present an abrupt transition to global 
synchronization.

\section{Clustered networks}

A small-world network has typically an average distance between 
nodes comparable to the value it would take on a purely random network, while
retaining an appreciable degree of clustering, as in regular networks.
Typical brain networks also display a small-world property \cite{hagmann}.
 
\begin{figure}[htbp]
\begin{center}
\includegraphics[height=7cm,width=8cm]{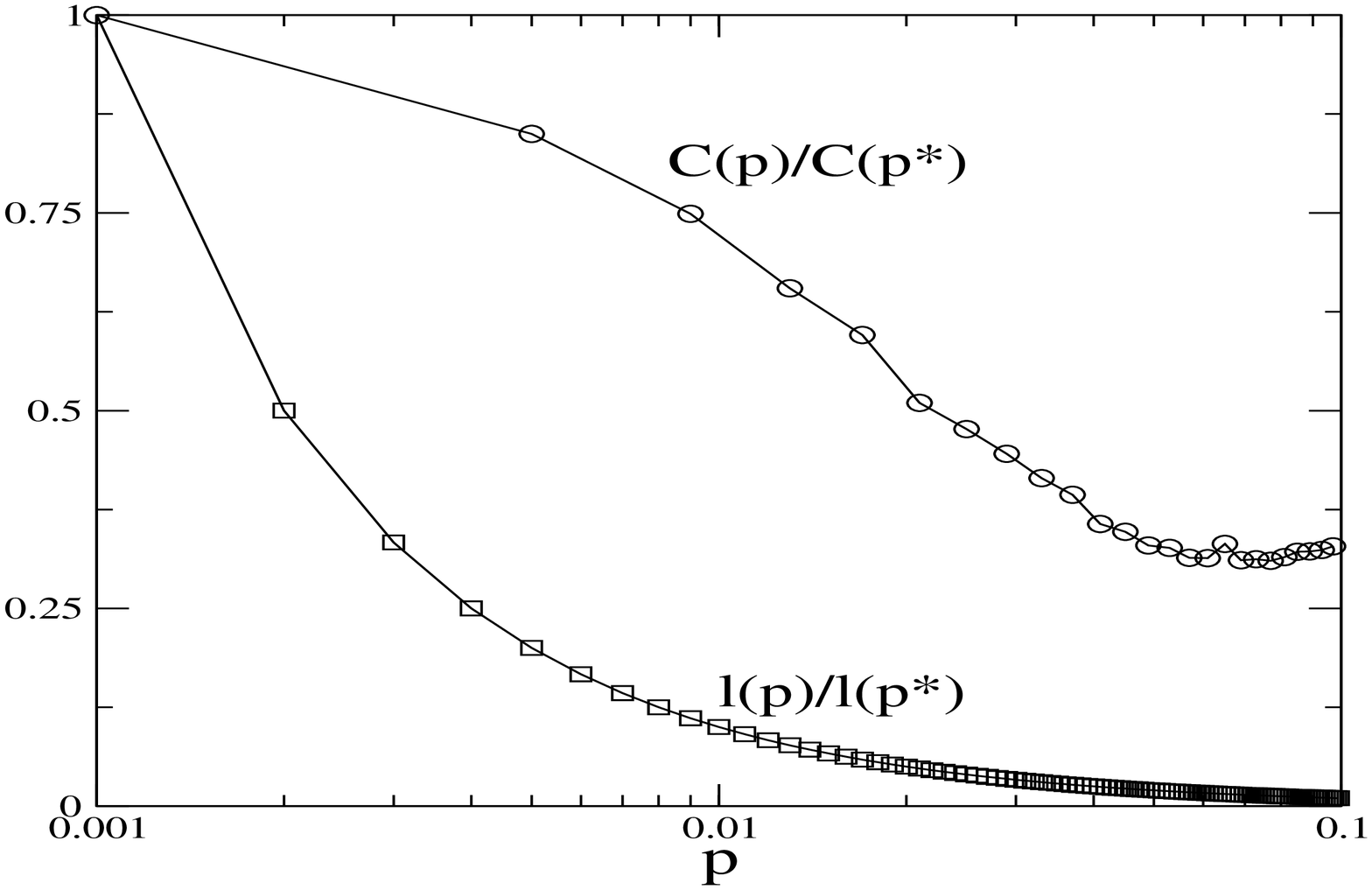}
\end{center}
\caption{Normalized clustering coefficient ($\circ$) and average
separation between node ($\square$) versus probability of non-local
connections, with $N=100$ and $p^*=10^{-3}$.}
\label{clusterl}
\end{figure}

We generate small-world networks following the procedure proposed 
by Newman and Watts \cite{newman}, inserting randomly chosen shortcuts in a 
regular network \cite{batista03}. To verify that the coupled map
network built according to procedure proposed by Newman and Watts has the
properties of a small-world network, we have computed the so-called
clustering coefficient, defined as the average fraction of pairs of 
neighbors of a site that happen also to be neighbors of each other, and the
average separation between sites, that they are showed in Fig. (\ref{clusterl})
by circles and squares, respectively, as a function of the probability.
The average network distance between sites must be of the same order as for
a random graph, on the other hand, the clustering coefficient must be much
greater than for a random graph. Then, in virtue of these requirements,
Fig. (\ref{clusterl}) suggests that the small-world property is displayed when
the value of the probability is around $10^{-2}$.

\begin{figure}[htbp]
\begin{center}
\includegraphics[height=4cm,width=5cm]{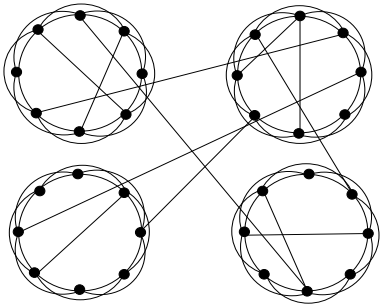}
\end{center}
\caption{Schematic representation of $M=4$ one-dimensional small-world
networks, each of $L=8$ nodes.}
\label{esq}
\end{figure}

We consider a network composed of  $M$ small-world subnetworks with $L$ neurons 
each (Fig. \ref{esq}). A given node has a probability $p_i$ to connect with 
other nodes inside the cluster, and a probability $p_o$ to be connected with 
neurons outside the cluster, the intercluster probability.

{\it Neuron model: } There are a number of mathematical models which emulate
neuronal activity, ranging from differential equations \cite{hindmarsh} to 
discrete-time maps \cite{rulkov}. We choose the Rulkov model 

\begin{eqnarray}
x (n+1) &=& f(x (n) ,y (n)) =  \frac{\alpha}{1 + [x (n)]^2} + y (n), \\
y (n+1) &=& g(x (n) , y (n)) = y (n) - \sigma x (n) - \beta,
\end{eqnarray}
where $x (n)$ is the fast and $y (n)$ is the slow dynamical variable. 
Furthermore, $\alpha$ affects directly the spiking time-scale and is chosen so that the time series of $x (n)$ presents an irregular sequence of spikes. The parameters $\sigma$ and $\beta$ describe 
the slow time-scale. 

{\it Phase Dynamics:} 
We consider the neurons to be non-identical, that is, they possess a 
mismatch in their internal parameters, we choose $\alpha$ as a mismatch parameter.  
Complete synchronization in networks of non-identical neurons is not possible. 
However, weaker synchronization such as phase synchronization 
\cite{pikovsky} may take place. 

To study phase synchronization, we need to introduce a phase for the chaotic 
dynamics of the bursts. This turns out to be a non straightforward task. For 
chaotic oscillators the phase cannot be defined unambiguous. Different ways 
to introduce a phase are possible, each one being chosen according to the 
particular case studied \cite{josi, pereira2}.

We consider the phase being increased
by $2 \pi $ at the beginning of each burst. We consider the burst to begin when 
the slow variable $y (n)$, which
presents nearly regular saw-teeth oscillations, has a local maximum, in
well-defined instants of time we call $t_k$. The duration of the chaotic
burst, $t_{k+1}-t_k$, depends on the variable $x (n)$ and fluctuates in an
irregular fashion as long as $x (n)$ undergoes chaotic evolution 
\cite{batista07}. 

We can define a phase describing the time evolution within 
each burst, varying linearly from $t_k$ to $t_{k+1}$
\begin{equation}
\phi (t)=2\pi k+2\pi \frac{t-t_k}{t_{k+1}-t_k}, \mbox{   }  t_k \le t < t_{k+1}
\end{equation}
where $t_k$ denotes the time occurrence of the $k$th burst.
The bursts occur in a coherent manner, that is, the time 
interval between burst have a small deviation. 

{\it Chemical Synapses:}
We analyze a neuron network coupled 
via excitatory synapse. We model the synapse as a static sigmoidal
nonlinear input-output function with a threshold and a saturation
parameters \cite{sompolinsky}. We order the neurons within the network from $1, 2, \ldots, ML$, and 
denote the dynamical variables of the $i$th neuron as $x_i(n)$ and $y_i(n)$.

The coupling function is given by  
$$ 
V( x_i( n), x_j (n)) = (x_i (n) -V_s) S( x_j( n)),
$$
where (the reversal potential) $V_s> x_i(n)$ for any $x_i (n)$ implies that the
synapse is excitatory. Here  $V_s = 2.0$. The synaptic coupling function $S$
is modeled by the sigmoidal function
\begin{equation}
S(x) = \frac{1}{1 + e^{-\lambda (x - \Theta_s)}}.
\end{equation}

We take the saturation parameter $\lambda = 10$, and   
the value of $\Theta_s = -0.25$
is chosen in such a way that every spike within a single neuron burst
can reach the threshold. 

The network dynamics is described by  
\begin{eqnarray}
x_i (n+1) & = &f(x_i (n),y_i (n))+ \varepsilon \sum_{j=1}^N a_{ij} V(x_i(n),x_j(n)), \nonumber \\
y_i(n+1) & = & g(x_i (n),y_i(n)),
\end{eqnarray}
where $i=1,2,...,ML$, 
$\sigma=\beta=0.001$, and $\alpha_i$ to be different for each node on uniform 
values deviate in the interval $[4.1,4.4]$. $\varepsilon>0$ is the overall 
coupling strength, and $A = (a_{ij})$ is the adjacency matrix, $a_{ij}$ is $1$ 
if neuron $i$ is connected to neuron $j$, and zero otherwise. 

The local part of the interaction considers the nearest and the
next-to-the-nearest neighbors of a given node. Moreover, with probability 
$p_i$ a new long range connection is created inside the each cluster, and with 
probability $p_o$ to create an outside connection.

\section{Global phase synchronization of bursting neurons}

The collective behavior of the network may be captured by the 
network mean field. 
$$
\langle x \rangle_{\mu}(n) = \frac{1}{ML}\sum_{i=1}^{ML} x_i(n).
$$
The emergence of a collective behavior enhances 
the mean field dynamics. 

\begin{figure}[htbp]
\begin{center}
\includegraphics[height=8cm,width=8cm]{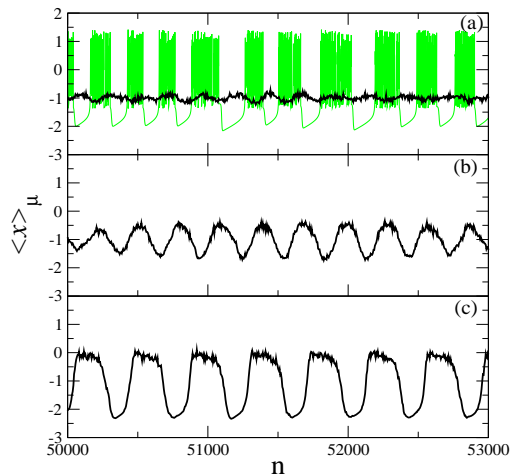}
\end{center}
\caption{Behavior of the mean field as a function of the overall coupling parameter
with fixed $p_i=10^{-2}$, $p_o=10^{-3}$, $M=4$ and $L=100$. 
a) $\varepsilon=0.01$ mean field presents 
no dynamics, but a noise-like behavior. b) $\varepsilon = 0.04$ some neurons 
are synchronized  and the mean field exhibit slows oscillators. c) 
$\varepsilon = 0.1$  close to global synchronization the mean field 
shows the bursting dynamics.}
\label{xn}
\end{figure}

To illustrate the mean field dependence on the synchronization 
we choose $M=4$ and $L=100$ with $p_i=10^{-2}$, $p_o=10^{-3}$.
For small values of the overall coupling parameter $\varepsilon=0.01$
the neurons burst in a nonsynchronized fashion. The mean field presents 
no dynamics, but a noise-like signal, fluctuating around the mean value Fig. \ref{xn}(a). 

The more synchronized neurons the higher the mean field. For a coupling 
$\varepsilon = 0.04$ the mean field already captures the dynamics of the 
burst (slow  time-scale) dynamics Fig. \ref{xn}(b).
For the chosen $p_i$ and $p_o$, if the coupling strength is large enough then 
the network globally burst phase synchronize. Close to global synchronization
$\varepsilon = 0.1$ 
the mean field already reveals the bursting dynamics, while the fast oscillation 
due to the spike are filtered, see Fig. \ref{xn}(c). This shows that even though the 
burst are synchronized the spikes can remain out of synchrony. 

{\it The order parameter} characterizes the emergence of synchrony in the network
\begin{equation}\label{porder}
z_n=R_n\exp (\rm{i}\Phi_n)\equiv \frac{1}{N}\sum_{j=1}^{N}\exp (\rm{ }i
\phi_n^{(j)}),
\end{equation}
where $R_n$ and $\Phi_n$ are the amplitude and angle, respectively, of a
centroid phase vector for a one-dimensional network with periodic boundary
conditions. If the bursting phases $\phi_n^{(j)}$ are spatially uncorrelated,
their contribution to the result of the summation in Eq. (\ref{porder}) is
small. However, in a globally phase synchronized state the
order parameter magnitude asymptotes the unity. 

The time averaged order parameter magnitude is given by
$$
{\bar R}= \frac{1}{T}\sum_{n=1}^T R_n,
$$
clearly depends on $\epsilon$ and both $p_i$, $p_o$ in and out
connection probability.  If the bursting dynamics is globally 
synchronized then $\overline{R} \approx 1$.

Our numerical analysis shows that clustered networks with good 
global synchronization properties is possible only for when a 
$p_o$ is larger than a certain critical value. This critical values
clearly depends on the number of subnetworks. 
For a network that contains many clusters a larger 
$p_o$ is required to achieve global synchronization.

\begin{figure}[htbp]
\begin{center}
\includegraphics[height=8cm,width=8cm]{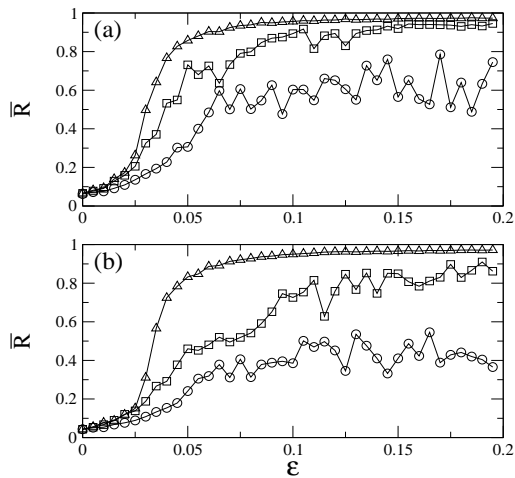}
\end{center}
\caption{Time averaged order parameter as a function of the 
coupling strength for $p_i=10^{-2}$ and $L=100$ fixed. a) $M=2$ and b) $M=4$. 
In both figures we have $p_o=0$ ($\circ$), $p_o=1.6\times 10^{-4}$ ($\square$), $p_o=2 \times 10^{-3}$ ($\triangle$).}
\label{pocluster}
\end{figure}

Fig. \ref{pocluster} shows the evolution of the average global order
parameter $\bar{R}$ for fixed $p_i=10^{-2} $ and $L=100$. 
In Fig. \ref{pocluster} a) we have $M=2$ and b) $M=4$, where we depict the 
behavior of $\overline{R}$ for various values of the $p_o$.
In the absence of outside connections and $\varepsilon$ large enough 
$\overline{R} \approx 1/M$, and if $p_o$ and $\varepsilon$ are large enough 
$\overline{R} \approx 1$. For $\varepsilon=0$ it must yield $\overline{R} 
\propto (ML)^{-1/2}$ so if $L \gg 1$ or $M \gg 1 $ $\overline{R}$ must be close
to zero.
 
The global synchronization is achieved when $\bar{R}=1$. 
However, if phase synchronization is achieved $\phi_i \approx \phi_j$ an almost 
global synchronized behavior can be obtain by setting a  threshold 
value near 1, we choose $\bar{R}=0.95$. The minimum coupling strength 
necessary to achieve $\bar{R}=0.95$ is the critical coupling 
strength $\varepsilon_c$, the critical coupling parameter for global burst phase synchronization.

{\it Phase Reduction:} the application of the Kuramoto phase reduction 
techniques \cite{kuramoto,liu} leads to an approximate phase equation to 
describe the collective phase dynamics of the neuron ensemble. Since the burst 
dynamics is coherent. It is possible to introduce a coordinate change such that
the phase of the i$th$ reads
$$
\phi_i(t) = 1 + \psi_i(t) + \zeta_i(t),
$$
where $\psi_i$ depends on the particular neuron and $\zeta$ is a small noise term \cite{josi}. 

Now, if the neurons are close to the onset of global synchronization 
$\phi_j \approx \phi_i$, the phase are described by $\psi_i$. Assuming that phase gradient is 
approximately constant along the trajectories, the phase 
reduction up to first order approximation renders
$$
\dot{\psi}_i(t) = \omega_i + b \sum_{j=1}^n a_{ij} (\psi_j - \psi_i),
$$
where $\omega_i$ depends on the neuron parameters and on the node degree.  
Close to the onset of synchronization $\psi_j - \psi_i \approx \sin(\psi_j - \psi_i)$. 
So as a result the phase dynamics of the neuron network can be described (up 
to first order approximation) by the Kuramoto model
$$
\dot\phi_i=\omega_i+\varepsilon\sum_{j=1}a_{ij}\sin(\phi_j-\phi_i),
$$
where, $\varepsilon$ is a rescaled overall coupling strength and $a_{ij}$ is the 
adjacency matrix. 

Recent results \cite{guan} have given bounds for the critical coupling 
strength required for global synchronization to the probabilities of 
intracluster and intercluster connections. The critical coupling strength reads
\begin{equation}\label{pred}
\varepsilon_c=\frac{C}{p_i (L-1)+p_o(N-L)},
\end{equation}
where the constant $C$ depends on the threshold value of the average global
order parameter $\bar{R}$ for defining global synchronization, that can be
regarded as a fitting parameter. Besides, a node is connected with $L-1$
nodes in the same cluster with probability $p_i$ that in our case is
$L-5$, and the node is connected with $N-L$ nodes from different cluster with
probability $p_o$. Thus, up to first order approximations, the onset of global bursting synchronization 
can be described in terms of Eq. (\ref{pred})

\begin{figure}[htbp]
\begin{center}
\includegraphics[height=8cm,width=8cm]{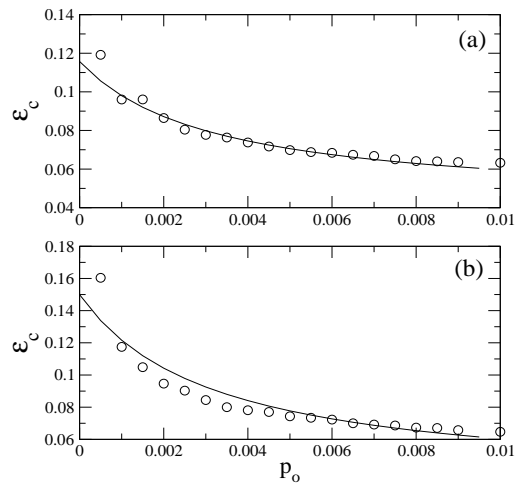}
\end{center}
\caption{$\varepsilon_c$ versus $p_o$ for a network with $L=100$ and
$p_i=10^{-2}$. The circles are for the threshold value of the global order 
parameter equal 0.95 and solid line 
$\varepsilon_c=C/[p_i(L-5)+p_o(N-L)]+\delta$, where (a) $M=2$, $C=0.07$ and 
$\delta=0.045$, and (b) $M=4$, $C=0.112$ and $\delta=0.032$.}
\label{critico}
\end{figure}

Our detailed numerical analysis corroborates the theoretical 
prediction. In Fig. \ref{critico} $p_i=10^{-2}$ and $L=100$, furthermore we use $M=2$ in Fig. a) and 
$M=4$ Fig. b). In both figures the circles represent the critical values in accordance with
the threshold global order parameter equal $0.95$. The solid curve is the theoretical prediction of the equation 
(\ref{pred}) with $\varepsilon_c+\delta$,

The probability $p_o$ increases and the critical coupling strength required for 
global synchronization decreases. This shows that as the network become are
homogeneous the global synchronization properties are enhanced. 

\section{Transition to phase synchrony}

The transition to synchronization occurs in an intermittent fashion, 
where $\bar R$ has laminar phases. To analyze the intermittent behavior  we introduce the quantity 
$$
F={1\over s}\sum_{1}^{s} N_R/\Delta_n,
$$ where 
$N_R$ is the number of occurrences of $\bar R>\xi$ ($\xi=0.95$) that we find 
within a time interval $\Delta_n$ and $s$ is the number of configurations of
the connections distributed with the determined probability $p_o$. 
Thus, $F$ 
can be interpreted as the fraction of global synchronization in a given time 
interval. 

\begin{figure}[htbp]
\begin{center}
\includegraphics[height=5cm,width=7cm]{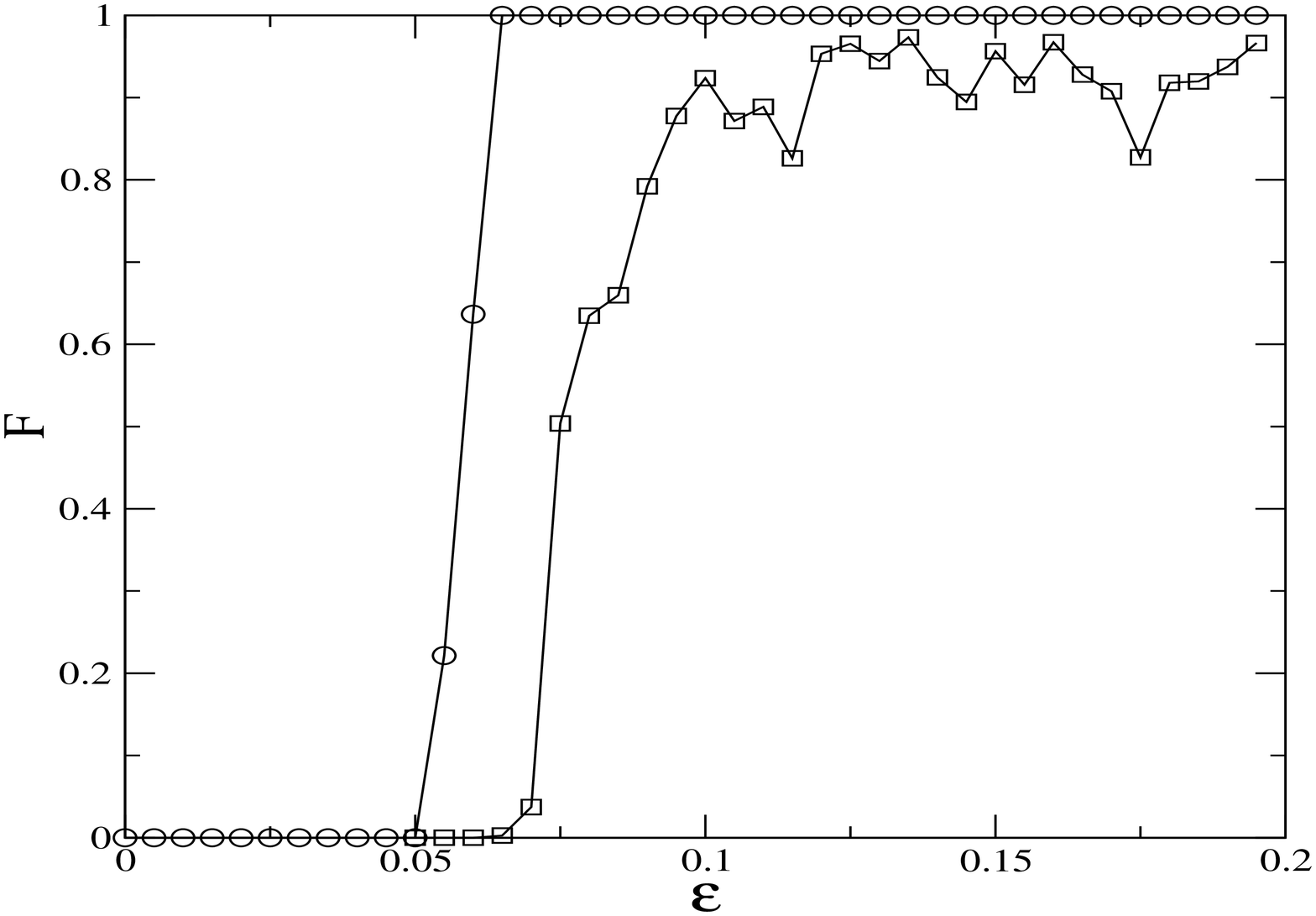}
\end{center}
\caption{ Fraction of time where the clustered network has larger 
${\bar R}$ than 0.95 {\it versus} coupling strength.
We consider a clustered network with $L=100$, $M=2$, $p_i=10^{-2}$,
$s=20$, and  $p_o=10^{-1}$ ($\circ$) and $p_o = 10^{-2}$ $(\square)$. This two 
probabilities leads to distinct transitions to global synchronization.}
\label{interm}
\end{figure}

$F$ depends on the coupling strength $\varepsilon$, and may present 
distinct behaviors as a function of the intercluster probability $p_o$.
In Fig. \ref{interm}(a) we depict the dependence of $F$ on $p_o$.
The transition to global synchronization is abrupt 
for $p_o=10^{-1}$ and $\varepsilon_c\approx 0.05$, while it is not abrupt for 
$p_o=10^{-2}$ and $p_o=2\times 10^{-3}$. 

Such behavior can be understood heuristically in terms of the synchronization 
properties of each isolated network. 
If $p_o$ is small enough, each network is almost independent. Hence, 
each network tend to synchronize by itself. So, as we increase the coupling 
each cluster synchronize and only then a behavior towards global synchronization is seen. 
This leads to a smooth transition. 
However, if $p_o$ is large then the collective dynamics of each subnetwork 
is strongly affected by the remaining subnetwork. So the synchronization 
occurs in one subnetwork is quickly spreads over the whole network, leading to 
a abrupt transition.

\section{Conclusions}

In conclusion, we presented a neuronal clustered network model in which the
neuron dynamical is described by the Rulkov map, furthermore the
connective architecture presents the small-world property. We demonstrated
the possibility of obtaining bursting global synchronization of Rulkov
neurons in a clustered networks, where the probability of intercluster
is varied and the  clusters exhibit small-world property. Moreover, the network
becomes more synchronizable when the probability of intercluster links is
increased. That is, for a fixed $p_i$ the increasing of $p_o$, the out connection probability, 
enhances global synchronization. This means that when heterogeneity between the communities
(clusters) is suppressed the network as a whole synchronizes better. 
 The critical coupling parameter required for  global synchronization as a function of the probability of intercluster 
presents the same behavior as Kuramoto-type dynamics.

\section*{Acknowledgments}
This work was made possible through partial financial support from the
following Brazilian research agencies: CNPq, CAPES and Funda\c c\~ao
Arauc\'aria.

\end{document}